\documentclass[conference]{IEEEtran}
\IEEEoverridecommandlockouts 

\usepackage{cite}
\usepackage{amsmath,amssymb,amsfonts}
\usepackage{graphicx}
\usepackage{dblfloatfix}
\usepackage{textcomp}
\usepackage{xcolor}
\usepackage{algorithm}
\usepackage{algpseudocode}
\usepackage{pgfplots}
\pgfplotsset{compat=1.18}
\usepackage{url}
\usepackage{booktabs}
\usepackage{enumitem}
\usepackage{xcolor}

\definecolor{headercolor}{RGB}{0,102,204}   
\definecolor{jsoncolor}{RGB}{153,0,0}

\def\BibTeX{{\rm B\kern-.05em{\sc i\kern-.025em b}\kern-.08em
		T\kern-.1667em\lower.7ex\hbox{E}\kern-.125emX}}

\begin{document}
\bstctlcite{BSTcontrol}

    \title{Beyond State Machines: Executing Network Procedures with Agentic Tool-Calling Sequences}

    \author{%
		\IEEEauthorblockN{%
			Purna Sai Garigipati\IEEEauthorrefmark{1}\IEEEauthorrefmark{2},
			Onur Ayan\IEEEauthorrefmark{1},
			Kishor Chandra Joshi\IEEEauthorrefmark{2},
			Xueli An\IEEEauthorrefmark{1}}
		\IEEEauthorblockA{\IEEEauthorrefmark{1}Heisenberg Research Center, Huawei Technologies Duesseldorf GmbH, 80992 Munich, Germany\\
			Email: \{purna.sai.garigipati,onur.ayan,xueli.an\}@huawei.com}
		\IEEEauthorblockA{\IEEEauthorrefmark{2}Eindhoven University of Technology, Eindhoven, The Netherlands\\
			Email: \{k.c.joshi\}@tue.nl}
	}
	
	\maketitle

\begin{abstract}
Agentic AI will be an essential enabling technology for designing future mobile communication systems, which could provide flexible and customized services, automate complex network operations, and drive autonomous decision-making across the network. This work studies how Large Language Model (LLM)-based network AI agents can be utilized to execute network procedures expressed as sequences of tool invocations. We investigate four approaches, which differ in how the agent obtains the procedure and in how execution is distributed between the agent and the underlying tools. We evaluated the latency and execution correctness across these approaches using a User Equipment (UE) IP allocation procedure as a case study. Furthermore, we conduct a stress test to examine how many sequential procedural steps an LLM agent can reliably execute before failure. Our results show that approaches relying on iterative agent-side reasoning incur higher latency and are more prone to execution errors, while approaches where the procedure is encapsulated within a single tool, which internally orchestrates the required steps by invoking other tools, reduce latency by limiting repeated reasoning. The stress-test results further show that the model with advanced tool-calling capability maintains reliable execution over longer procedures than the other evaluated models; however, all models exhibit reliability degradation as procedure length increases, revealing clear execution limits in multi-step tool-based workflows. To systematically analyze failures in procedure execution, we introduce a procedure-specific error taxonomy that categorizes deviations in multi-step procedural execution.
\end{abstract}


\section{Introduction}
\label{sec:introduction}

Agents empowered by Large Language Models (LLMs) have introduced a new paradigm of autonomous systems capable of reasoning, planning, and interacting with external tools to accomplish complex tasks. Such systems extend beyond single-shot inference and operate through iterative decision-making and tool invocation. This paradigm shift has gained significant attention because it enables complex workflows to be executed without explicitly hard‑coded control logic, instead relying on the model inference to determine the sequence of actions required to achieve a given objective.

Recent work shows that agentic approaches are being actively explored in the context of next-generation networks, particularly in 6G. Several studies focus on the Radio Access Network (RAN), where agentic frameworks have been proposed for real-time control, resource management, and optimization \cite{navidan2026autonomousoranmultiscaleagentic, bandara2026agenticaicontrolplane,feng20256gnativeaiedgenetworks}. In addition, recent efforts consider end-to-end intelligence across both RAN and core networks, integrating monitoring, policy control, and cross-layer optimization using agentic approaches \cite{han2026e2eintelligence6gnetworks, jiang2026agenticaiempoweredintentbased, xiao2025sanetsemanticawareagenticai}. Collectively, these works indicate a shift toward agent-driven network automation.

In this context, we study the use of LLM-based agents to execute network procedures. Typically, a network procedure consists of a sequence of dependent operations that must be executed in a strict order to achieve a target system state. Traditionally, such procedures are implemented as scripts or workflows, which require explicit development, testing, and deployment. Supporting variability in network conditions often requires extensive conditional logic, making these implementations difficult to scale and maintain. Moreover, such implementations are tightly coupled to predefined states and can fail when the observed network state deviates from expected conditions.

In contrast, an agent-based approach enables procedures to be specified at a higher level using natural language descriptions. Given a set of standardized tools, different procedures can be dynamically composed by varying the sequence of tool invocations based on the task and intermediate outcomes, without requiring explicit reprogramming. This flexibility is important in scenarios where a network operator or an application needs to execute a new procedure that is not predefined in existing specifications.

This raises a fundamental question: can an LLM-based agent reliably execute telecom-grade procedures? To answer this, we study the agent’s ability to perform sequential tool invocation, focusing on whether the correct step ordering is maintained, how the execution behaves across repeated runs, what types of procedural violations occur, and how the reliability changes as the length of the procedure increases. We further analyze different mechanisms for providing the procedure to the agent, reflecting realistic deployment scenarios.

Although agentic systems enable flexible execution, previous work shows that LLM-based agents are prone to failures during multi-step reasoning and tool interaction, and existing studies have proposed taxonomies to characterize such failures, covering aspects such as planning errors, tool invocation issues, and execution inconsistencies \cite{zhu2025llmagentsfaillearn, shah2026characterizingfaultsagenticai, 11081716}. However, these taxonomies are designed for open-ended or generalized domains and do not address the strict constraints of sequential tool execution. We therefore define a procedure-specific error taxonomy tailored to this setting, enabling precise analysis of procedural execution correctness for agentic systems.

The main contributions of this paper are summarized as follows:
\begin{itemize}
    \item \textbf{Procedural Execution Approaches:} We present four approaches for delivering procedural logic to LLM agents and characterize how procedure placement impacts end-to-end latency and execution correctness.
    
    \item \textbf{Scalability Limits of Sequential Tool Execution:} We identify the execution limits of LLM agents by showing how performance degrades as the number of sequential steps increases.

    \item \textbf{Procedure-Specific Error Taxonomy:} We define an error taxonomy tailored to sequential tool execution, enabling precise classification of failures in multi-step agentic workflows.

\end{itemize}

The remainder of this paper is organized as follows. Section~\ref{sec:methodology} formulates the procedural execution model, defines the evaluation metrics, and presents the execution approaches and error taxonomy. Section~\ref{sec:experimental_evaluation} describes the experimental scenarios and presents the evaluation results. Section~\ref{sec:Conclusion} concludes the paper.

\section{Methodology}
\label{sec:methodology}

This section presents the procedural execution model considered in this work, the performance metrics used for evaluation, the four execution approaches, and the error taxonomy adopted for analyzing failures.

\subsection{Definition of a Procedure as Tool Sequence} 
\label{subsec:system_model}

We consider a task execution setting in which an LLM-based agent interacts with a tool server that exposes a total number of $m$ tools, denoted by $\mathcal{T} = \{\tau_1, \tau_2, \dots, \tau_m\}$. Given a user intent $i$, the agent must identify and execute the matching procedure $P_i$ with $P_i \in \mathcal{P}$ where $\mathcal{P}$ denotes the set of procedures available at the agent. We define procedure $P_i$ as an ordered sequence of tool calls, where each required tool $\tau_{i,j}$ belongs to the available toolset $\mathcal{T}$:
\begin{equation}
P_i = (\tau_{i,1}, \tau_{i,2}, \dots, \tau_{i,k}),
\label{eq:procedure}
\end{equation}

Here, the index $i$ represents the procedure that matches the user's intent from a set of possible procedures $\mathcal{P}$. The second index represents the step number  within the procedure. For example, $\tau_{i,1}$ is the first tool occurring in the procedure $P_i$ followed by $\tau_{i,2}$ up to the last tool $\tau_{i, k}$.

The observed execution produced by the agent is similarly represented as an ordered sequence of tool calls:
\begin{equation}
O_i = (\hat{\tau}_{i,1}, \hat{\tau}_{i,2}, \dots, \hat{\tau}_{i,\hat{k}}),
\label{eq:observed}
\end{equation}
where $\hat{\tau}_{i,j}$ denotes the tool invoked at step $j$ during execution of procedure $i$, and $\hat{k}$ is the number of executed steps.

In this formulation, $P_i$ represents the correct procedure (i.e., ground-truth), while $O_i$ denotes the sequence of tool calls executed by the agent.

\subsection{Evaluation Metrics}
\label{subsec:metrics}

We evaluate procedural execution using latency and execution correctness. The total latency cost $C(i)$ is defined as the end-to-end execution time required to process intent $i$, including both the LLM reasoning and tool invocation:
\begin{equation}
C(i) = \sum_{j=1}^{N_{\text{llm}}} L_j^{\text{llm}} + \sum_{j=1}^{\hat{k}} L_j^{\text{tool}},
\label{eq:cost}
\end{equation}
where $N_{\text{llm}}$ is the number of LLM reasoning steps, $\hat{k}$ is the number of tool invocations executed, and $L_j^{\text{llm}}$ and $L_j^{\text{tool}}$ denote the latency of the corresponding reasoning and tool-execution step, respectively.\footnote{Transmission latency is negligible and is therefore excluded from the latency cost equation.} 

Execution correctness is measured using a binary reliability metric $R(P_i)$. For a single run, it evaluates whether the observed sequence of actions $O_i$ perfectly matches the expected procedure $P_i$:
\begin{equation}
R(P_i) = \begin{cases} 
0, & \text{if } k \neq \hat{k} \text{ or } \tau_{i, j} \neq \hat{\tau}_{i, j} \text{ for at least one } j \\ 
1, & \text{otherwise} 
\end{cases}
\label{eq:reliability}
\end{equation}

By this definition, a score of 0 indicates that the observed sequence $O_i$ differs from $P_i$ in length ($k \neq \hat{k}$) or composition ($\tau_{i, j} \neq \hat{\tau}_{i, j}$), which implies an erroneous procedure execution as categorized later in Section~\ref{subsec:error_taxonomy}. Averaging $R(P_i)$ over repeated runs yields the execution correctness rate for a given model and approach.

\begin{figure*}[!t]
    \centering
    \includegraphics[width=0.95\linewidth]{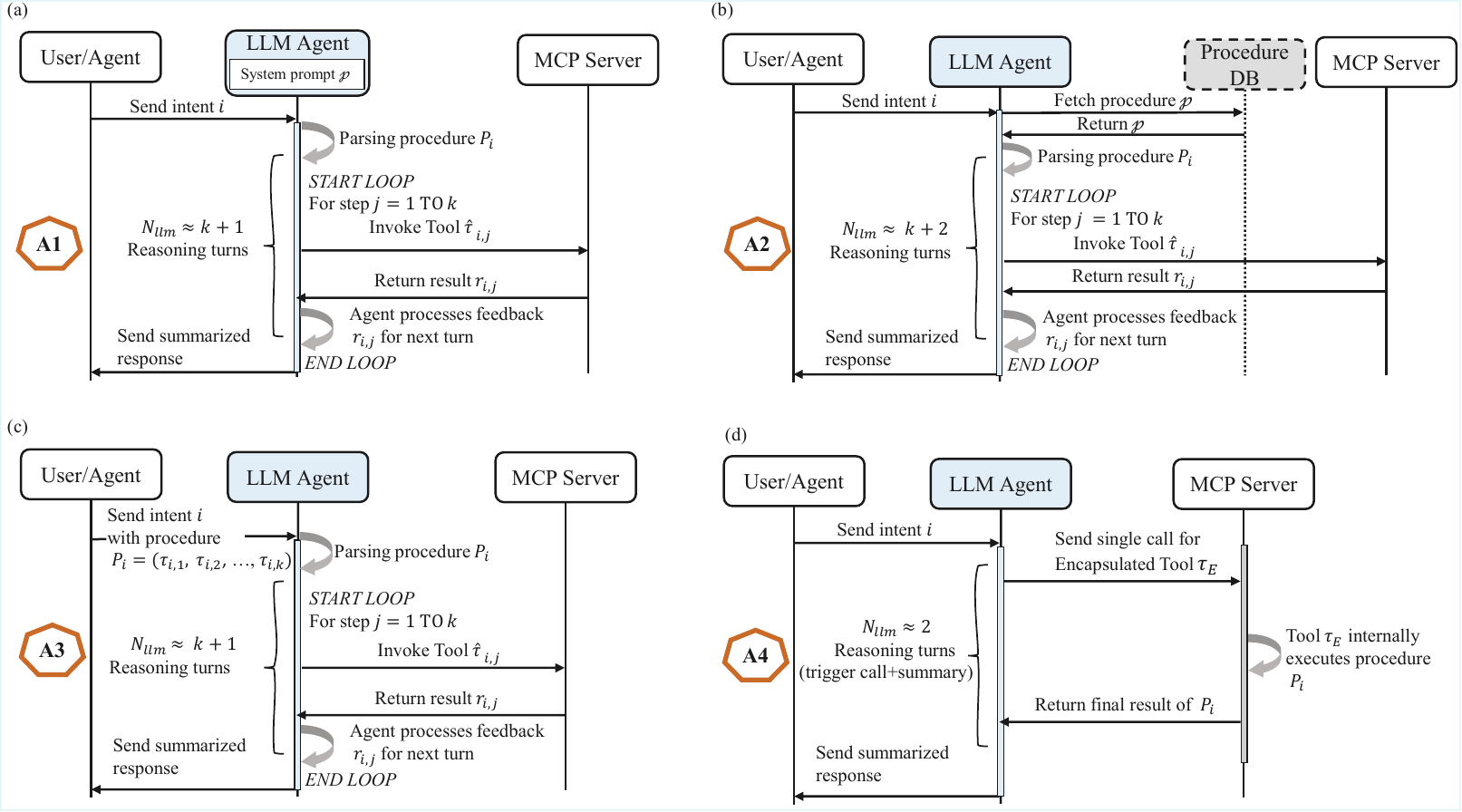}
    \caption{Comparison of four procedural execution approaches. (a) A1 embeds the procedure within the agent, (b) A2 retrieves the procedure from an external database, (c) A3 receives the procedure in the input prompt, and (d) A4 encapsulates the procedure within a single tool. The figure highlights the difference between iterative multi-step execution (A1--A3) and single-call execution (A4).}
    \label{fig:approaches}
\end{figure*}

\subsection{Procedural Execution Approaches}
\label{subsec:approaches}

The four approaches considered in this work are shown in Fig.~\ref{fig:approaches}. They share the same basic agent--tool interaction setting, but differ in where the procedure is defined and how the sequential execution is carried out, particularly in the distinction between iterative agent-driven execution and tool-encapsulated execution.

\begin{itemize}
    
    \item \textbf{A1: Agent-Embedded Procedure:} The set of procedures $\mathcal{P}$ is available to the agent as its system prompt. After receiving intent $i$, the agent must first parse the prompt to identify and extract the matching procedure $P_i \in \mathcal{P}$. Once identified, the agent reasons over this specific sequence and invokes the required tools step by step. For a procedure of length $k$, this leads to approximately $N_{\text{llm}} \approx k + 1$ reasoning steps,\footnote{Values are approximate as they represent the optimal execution path; in practice, model errors or tool failures often cause the agent to deviate, leading to additional reasoning turns or early termination.} where the final step corresponds to response summarization.
    \item \textbf{A2: Server-Provided Procedure:} The agent first retrieves the procedure information $\mathcal{P}$ from an external database or repository, parses the specific procedure $P_i$ for the intent, and then executes it sequentially in the same manner as A1. This additional retrieval step increases the number of reasoning steps to $N_{\text{llm}} \approx k + 2$.

    \item \textbf{A3: User/Agent-Provided Procedure:} The specific procedure $P_i$ is explicitly specified within the incoming prompt, which may originate from a user or another agent. The LLM agent parses the sequence and executes it iteratively, resulting in a reasoning overhead comparable to A1, i.e., $N_{\text{llm}} \approx k + 1$.

    \item \textbf{A4: Tool-Encapsulated Procedure:} The entire logic of $P_i$ is encapsulated inside a single tool ($\tau_E$). After receiving the user intent, the agent selects and invokes this tool once, and the internal procedure is executed within the tool implementation. Thus, the number of LLM reasoning steps is reduced to approximately $N_{\text{llm}} \approx 2$, corresponding to one trigger call and one final summarization step. This shifts execution complexity from the LLM to deterministic tool logic.
\end{itemize}

\subsection{Error Taxonomy for Tool-Based Procedures}
\label{subsec:error_taxonomy}

Whenever a run fails, i.e., when $R(P_i)=0$, the observed deviation is categorized to identify the failure mode. We adopt a strict error taxonomy because exact procedural execution is essential in structured network tasks.

\begin{figure*}[!t]
    \centering
    
    \includegraphics[width=.9\linewidth]{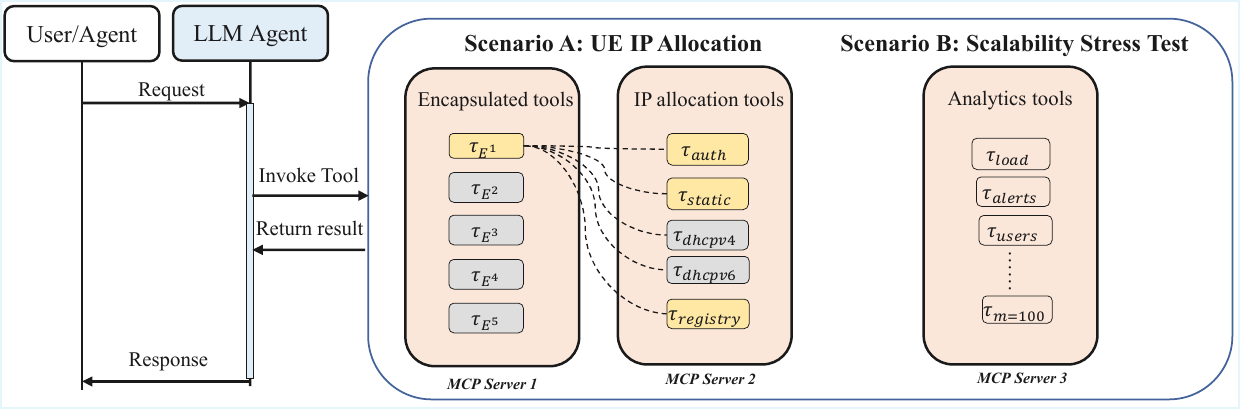}
    \caption{Overview of the experimental setups. Scenario A illustrates the UE IP Allocation workflow across two MCP servers: MCP Server 2 provides the IP allocation tools used in approaches A1--A3, while MCP Server 1 hosts encapsulated tools used in approach A4. The highlighted tools in Scenario A indicate the subset of tools executed for the representative request. Scenario B utilizes MCP Server 3, providing a pool of $m=100$ network analytics tools to stress-test sequence stability for procedures of length up to $k=50$.}
    \label{fig:experimental_setup}
\end{figure*}

\begin{enumerate}
\item \textbf{Wrong Tool:} A \emph{Wrong Tool} error occurs when the agent fails to correctly invoke the required tool for a given step in the procedure. This can manifest in three forms:
\begin{itemize}[noitemsep, topsep=2pt, leftmargin=*]
    \item \textbf{Tool Outside Procedure:} The agent invokes a tool $\hat{\tau}_{i,j}$ that does not appear in $P_i$, i.e., a tool that does not belong to the intended procedure.
    
    \item \textbf{Wrong Tool Name:} The agent intends to use the correct tool but invokes it using an incorrect or hallucinated name.
    
    \item \textbf{Wrong Parameters:} The agent invokes the correct tool but provides incorrect, missing, or invalid input arguments.
    
\end{itemize}
This corresponds to incorrect tool invocation and represents a critical deviation, as it introduces actions outside the intended execution logic. Due to this severity, any execution containing both \emph{Wrong Tool} and \emph{Duplicate Tool} errors is strictly classified as a \emph{Wrong Tool} error.
    
    \item \textbf{Duplicate Tool:} A \emph{Duplicate Tool} error is recorded when a valid tool $\tau \in P_i$ is invoked multiple times unnecessarily within $O_i$, without progressing the execution state. This typically reflects a reasoning loop and leads to an increase in both the number of reasoning steps $N_{\text{llm}}$ and, subsequently, the latency cost $C(i)$.

    \item \textbf{Premature Stop:} A \emph{Premature Stop} error occurs when the execution terminates before completing all $k$ steps of the procedure, even though the executed tools are in the correct order up to the stopping point. In other words, $O_i$ is a proper prefix of $P_i$ with  $\hat{k} < k$.

    \item \textbf{Wrong Order:} A \emph{Wrong Order} error is recorded when all required tools are present in $O_i$, but the sequence does not match the prescribed ordering of $P_i$, i.e., $O_i \neq P_i$ despite containing the same elements.

    \item \textbf{No Tool Calls:} A \emph{No Tool Calls} error is defined when the observed sequence is empty, i.e., $O_i = \emptyset$, indicating that no tool invocation was performed despite the requirement for procedural execution.
\end{enumerate}

\section{Experimental Evaluation}
\label{sec:experimental_evaluation}

To evaluate the performance and stability of LLM-based procedure execution, we designed two experimental scenarios, as shown in Fig.~\ref{fig:experimental_setup}. Scenario A compares the four approaches introduced in Section~\ref{subsec:approaches}, while Scenario B serves as a stress test to evaluate how the correctness of sequential tool execution is affected by the target procedure length $k$.

\subsection{Scenario A: UE IP Allocation}
\label{subsec:telecom_setup}

Scenario A simulates a simplified User Equipment (UE) IP allocation procedure. At a high level, the procedure first authorizes the UE and the requested session type, then checks whether a static IP is already assigned. If no static IP for the given UE is found, the network allocates an IP address using the appropriate DHCP server. Finally, the assigned IP is recorded in the registry. We adapted the logical guidelines of 3GPP standards~\cite{3gpp23501, 3gpp24501} into a set of executable network tools, where the orchestration procedure is executed through LLM-driven tool invocation.

As shown in Scenario A of Fig.~\ref{fig:experimental_setup}, MCP Server 2 hosts the IP allocation tools used directly in approaches A1, A2, and A3, where the agent executes the procedure step-by-step. These tools are defined as follows:
\begin{enumerate}
    \item \textbf{UE Authorization ($\tau_{auth}$):} Validates the UE ID and the requested session type.
    \item \textbf{Static IP Retrieval ($\tau_{static}$):} Checks whether a static IP is pre-assigned for the UE.
    \item \textbf{Dynamic IPv4 Allocation ($\tau_{dhcpv4}$):} Dynamically allocates an IPv4 address.
    \item \textbf{Dynamic IPv6 Allocation ($\tau_{dhcpv6}$):} Dynamically allocates an IPv6 address.
    \item \textbf{IP Assignment ($\tau_{registry}$):} Finalizes IP assignment by notifying the UE and recording the allocation in the network registry.
\end{enumerate}

The agent is provided with procedure information through the approaches described in Section~\ref{subsec:approaches}. In this setup, the procedures cover IP address allocation for IP PDU sessions, including IPv4, IPv6, and IPv4v6 cases. The correct tool-call sequence depends on both the request and the outputs of intermediate tool calls.

For example, consider an IPv4 address allocation request for a UE. The first step is authorization. If authorization fails, the correct procedure has length $k=1$:
\[
P_i = (\tau_{auth}).
\]
If the UE is authorized, the next step checks whether a static IP address is configured. In Scenario A, the evaluated request corresponds to a UE with an available static IP address. Therefore, the correct procedure has length $k=3$:
\[
P_i = (\tau_{auth}, \tau_{static}, \tau_{registry}).
\]
That is, the agent must authorize the UE, retrieve the static IP address, skip the dynamic IP address allocation steps (i.e., $\tau_{dhcpv4}$ and $\tau_{dhcpv6}$), and finalize the IP assignment. Otherwise, if no static IP address is available, the procedure continues with the appropriate DHCPv4 or DHCPv6 tool before finalization. Thus, for approaches A1, A2, and A3, intermediate reasoning is required because the agent must use each tool output to determine the next action.

\begin{figure*}[!t]
    \centering
    \includegraphics[width=.9\linewidth]{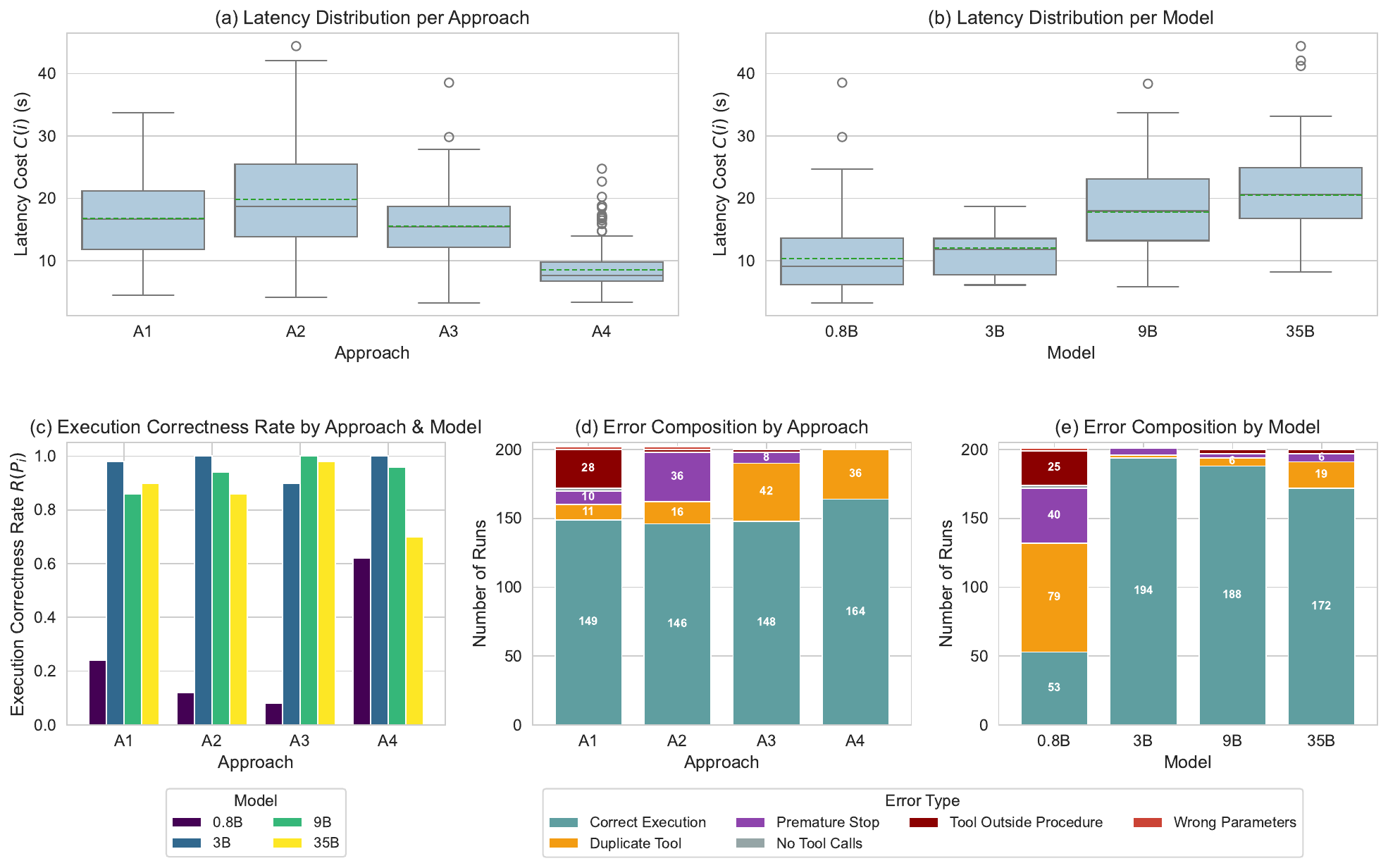}
    \caption{Evaluation of the UE IP Allocation procedure (Scenario A). The top row displays the end-to-end latency cost $C(i)$ categorized by (a) approach and (b) model size. The bottom row illustrates (c) the execution correctness rate, representing the average of the reliability metric $R(P_i)$, alongside the distribution of specific error types across (d) approaches and (e) models. Note that 0.8B, 3B, 9B, and 35B refer to the Qwen-0.8B, Qwen-Coder-3B, Qwen-9B, and Qwen-35B models, respectively.}
\label{fig:scenario_a_results}
\end{figure*}

MCP Server 1 hosts a set of encapsulated tools used in approach A4. Each encapsulated tool corresponds to a particular procedure and deterministically invokes the required lower-level tools that constitute the procedure one after another. In this setup, $\tau_{E^1}$ corresponds to the UE IP allocation procedure, while the other encapsulated tools correspond to different procedures. Thus, in A4, the agent no longer has to invoke multiple tools to perform $P_i$. Instead, it only selects one of the encapsulated tools located at the MCP Server 1 and subsequently invokes it. Note that this implies a single inference step to identify the right tool at the beginning, while in A1, A2, and A3 multiple inference steps during the procedure execution are necessary.

The nature of the input intent $i$ varies by approach. For A1, A2, and A4, the intent specifies only the desired outcome and the request parameters, such as the target \texttt{ue\_id} and \texttt{session\_type}, without explicitly providing the tool-call sequence. For A3, the input is explicit, where we assume that a client-side agent associated with the user provides the procedure $P_i$ directly in the request, i.e., the required tools and their order are specified as part of the input.

We evaluated four LLMs to assess the impact of model scale and tool-calling capability:
\begin{itemize}
    \item \textbf{Qwen 3.5:} Models evaluated at different parameter scales (0.8B, 9B, and 35B) to study the impact of model size on sequence stability, hereafter referred to as \textit{Qwen-0.8B}, \textit{Qwen-9B}, and \textit{Qwen-35B}.
    \item \textbf{Qwen3-Coder-Next:} A model with advanced tool-calling capability and approximately 3B active parameters per token during inference, hereafter referred to as \textit{Qwen-Coder-3B}.
\end{itemize}

To ensure statistical significance, the IPv4 allocation request was executed $50$ times per model for each approach, yielding $200$ independent runs per approach in Scenario A. The results for Scenario A are summarized in Fig.~\ref{fig:scenario_a_results}.

\textbf{Latency Cost:} Fig.~\ref{fig:scenario_a_results}(a) shows that latency $C(i)$, defined in (\ref{eq:cost}), varies across approaches primarily due to differences in the number of reasoning turns. Approaches A1 and A3 exhibit similar latency, as both require comparable multi-step reasoning. Approach A2 incurs the highest latency due to the additional procedure retrieval step. In contrast, Approach A4 achieves the lowest latency by reducing LLM interaction to a single trigger call. From a model perspective, Fig.~\ref{fig:scenario_a_results}(b) indicates that latency increases with model size, with Qwen-35B exhibiting the highest delay, while Qwen-0.8B shows the lowest latency but a wider spread, which is consistent with the unstable executions discussed below.

\textbf{Execution Correctness and Errors:} Execution correctness $R(P_i)$ across approaches and models is shown in Fig.~\ref{fig:scenario_a_results}(c). For this short procedure ($k=3$), Qwen-Coder-3B, Qwen-9B, and Qwen-35B consistently achieve high execution correctness rates, while Qwen-0.8B performs poorly.

The distribution of error types further explains these outcomes. Across approaches, Fig.~\ref{fig:scenario_a_results}(d) shows that A4 achieves the best overall approach-level performance, with the highest number of correct executions and the lowest number of errors. This indicates that encapsulating the procedure inside a deterministic tool reduces step-by-step execution failures. Errors in A4 are mainly due to \emph{Duplicate Tool} errors, which occur primarily for the Qwen-0.8B model and indicate repeated invocation of the encapsulated tool rather than failures in the internal execution of the procedure. Fig.~\ref{fig:scenario_a_results}(e) further shows that most errors occur for Qwen-0.8B, which frequently exhibits \emph{Duplicate Tool}, \emph{Premature Stop}, and \emph{Tool Outside Procedure} errors. In contrast, Qwen-Coder-3B achieves the highest number of correct executions and provides the best trade-off between latency and correctness. Qwen-9B and Qwen-35B also perform well for this short procedure, although Qwen-35B shows more errors than Qwen-9B despite its larger size. Finally, we observe that \emph{Wrong Order} errors and \emph{Wrong Tool Name} errors do not occur in any of the evaluated configurations.

\begin{figure}[!t]
    \centering
    \includegraphics[width=0.94\linewidth]{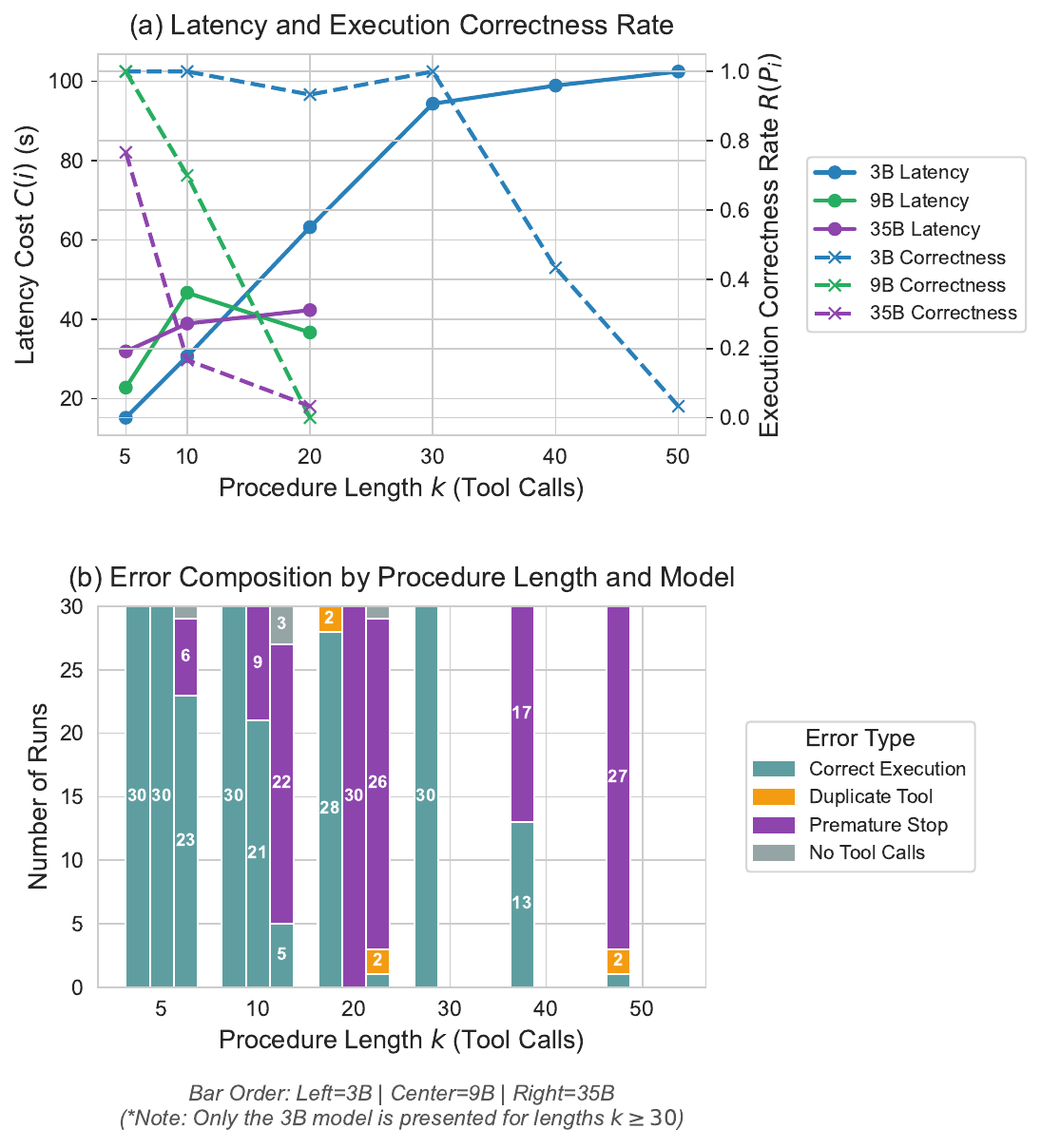}
    \caption{Scalability stress test results (Scenario B). Panel (a) illustrates the trade-off between mean latency cost $C(i)$ and execution correctness rate $R(P_i)$ as the target procedure length $k$ increases. Panel (b) details the composition of procedural errors at varying procedure lengths. Note that 3B, 9B, and 35B refer to the Qwen-Coder-3B, Qwen-9B, and Qwen-35B models, respectively.}
\label{fig:scenario_b_results}
\end{figure}

\subsection{Scenario B: Scalability Stress Test}
\label{subsec:stress_test_setup}

Because Scenario A requires only a short tool sequence ($k=3$), it does not fully expose the limitations of long multi-step reasoning loops. To identify the operational breaking point of these LLMs, we introduced a sequential stress-test setup using MCP Server 3, as shown in Scenario B of Fig.~\ref{fig:experimental_setup}.

Using approach A1, we expanded the tool pool $\mathcal{T}$ inside MCP Server 3 to contain $m=100$ network analytics tools (e.g., Average Cell Throughput, AMF Load, and Handover Success Rate). Each tool accepts a target geographic region as input and returns the corresponding KPI. We define the user intent $i$ as a query to analyze the network health of a target region. To fulfill this request, the agent must execute target procedures $P_i$ of varying lengths $k \in \{5, 10, 20, 30, 40, 50\}$.

For a given length $k$, the agent must sequentially invoke $k$ distinct tools to gather the required KPIs, summarize the collected data, and flag any abnormal metrics. This isolates and stresses the multi-turn reasoning loop $N_{\text{llm}}$, allowing us to evaluate how well the models maintain sequence stability as the procedure length increases. We performed 30 independent runs per procedure length for each evaluated model. The 0.8B model was excluded from this phase, as the results from Scenario A showed that its capabilities were insufficient for extended logic loops. The results for Scenario B are summarized in Fig.~\ref{fig:scenario_b_results}.

\textbf{Sequence Scalability:} Fig.~\ref{fig:scenario_b_results}(a) shows how execution correctness rate and latency evolve as the procedure length $k$ increases. Latency increases with sequence length, reflecting the growing number of reasoning steps required for longer procedures. Qwen-9B and Qwen-35B degrade rapidly with increasing $k$, with execution correctness rates dropping sharply by $k=20$. Notably, Qwen-35B exhibits a faster decline than Qwen-9B, indicating that larger parameter size does not necessarily translate to improved robustness in long sequential reasoning tasks. In contrast, Qwen-Coder-3B maintains near-perfect execution correctness up to $k=30$.

\textbf{Breaking Point and Failure Analysis:} The error composition in Fig.~\ref{fig:scenario_b_results}(b) explains the observed degradation. As $k$ increases, failures are dominated by \emph{Premature Stop} errors, indicating that the agent terminates execution before completing the full sequence. This trend is consistent across models but occurs significantly earlier for Qwen-9B and Qwen-35B. Beyond $k=30$, even Qwen-Coder-3B, which has advanced tool-calling capability, begins to degrade, with execution correctness dropping substantially by $k=50$, indicating a practical upper bound on reliable multi-step execution where maintaining consistency across long interaction histories becomes challenging.

\section{Conclusion}
\label{sec:Conclusion}

In this work, we investigated how LLM-based agents execute network procedures through sequential tool invocations. We compared four execution approaches and showed that approaches relying on iterative agent-side reasoning incur higher latency and are more error-prone, while the tool-encapsulated approach achieves lower latency and higher execution correctness by reducing repeated reasoning. Across the evaluations, we also observed that increasing model size alone does not necessarily improve robustness in tool-calling sequences. Finally, the stress-test results showed that the model with advanced tool-calling capability maintains reliable execution over longer procedures than the other evaluated models; however, all models ultimately degrade as the number of sequential tool calls increases, revealing clear breaking points in long procedure execution.

As future work, we plan to fine-tune a base model using execution traces labeled with our error taxonomy to evaluate whether this improves long sequential tool execution. We also plan to investigate mechanisms such as agent harness and reusable agent skills to further stabilize complex telecom-grade network procedures.

\section*{Acknowledgements}
This work has been supported by the European Union’s Horizon Europe MSCA-DN programme through the SCION Project under Grant Agreement No. 101072375.

\bibliography{bibfile}
		
\end{document}